\documentclass[letterpaper]{article} 
\usepackage{aaai2026}  
\usepackage{times}  
\usepackage{helvet}  
\usepackage{courier}  
\usepackage[hyphens]{url}  
\usepackage{graphicx} 
\usepackage{natbib}  
\usepackage{caption} 
\DeclareCaptionStyle{ruled}{labelfont=normalfont,labelsep=colon,strut=off} 
\usepackage{algorithm}
\usepackage{algorithmic}
\usepackage{color, colortbl}
\usepackage[table]{xcolor}
\usepackage{url}
\usepackage{subcaption}
\usepackage{textcomp}
\usepackage{soul}
\usepackage{multirow}
\usepackage{enumitem}
\usepackage{mathtools}
\usepackage{siunitx}

\usepackage{booktabs} 
\usepackage{longtable,booktabs}

\usepackage{arydshln}
\definecolor{linkColor}{RGB}{6,125,233}
\definecolor{green}{rgb}{0.0, 0.65, 0.31}
\definecolor{bleudefrance}{rgb}{0.19, 0.55, 0.91}
\definecolor{ceruleanblue}{rgb}{0.16, 0.32, 0.75}
\definecolor{grey}{HTML}{969696}
\definecolor{violet}{HTML}{756bb1}
\definecolor{dgrey}{HTML}{01665e}
\definecolor{lgrey}{HTML}{5ab4ac}
\definecolor{dgreen}{HTML}{005a32}
\definecolor{purple}{HTML}{ae017e}

\definecolor{editCol}{HTML}{000000}
\definecolor{maskCol}{HTML}{c51b7d}
\definecolor{lrColor}{HTML}{8856a7}
\definecolor{trColor}{HTML}{d01c8b}
\definecolor{ctColor}{HTML}{4dac26}
\definecolor{brickred}{HTML}{f03b20}

\definecolor{sigone}{RGB}{222, 235, 247}    
\definecolor{sigtwo}{RGB}{158, 202, 225}    
\definecolor{sigthree}{RGB}{107, 174, 214}  

\definecolor{improveCol}{HTML}{4dac26}
\definecolor{worsenCol}{HTML}{d01c8b}

\definecolor{DarkBlue}{HTML}{00008B}
\definecolor{mscolor}{HTML}{01665e}
\definecolor{nmscolor}{HTML}{bf812d}
\definecolor{lgreen}{HTML}{ccece6}
\definecolor{dolive}{HTML}{308014}

\colorlet{tablerowcolor4}{gray!50} 

\newcommand*{\textlabel}[2]{%
  \edef\@currentlabel{#1}
  \phantomsection
  #1\label{#2}
}

\colorlet{tableheadcolor}{gray!25} 
\colorlet{tablerowcolor}{gray!10} 
\colorlet{tablerowcolor2}{gray!45} 
\colorlet{tablerowcolor3}{gray!12} 

\newcommand{\rowcolmedium}{\rowcolor{tablerowcolor2}}
\newcommand{\rowcollight}{\rowcolor{tablerowcolor3}} %

\newcolumntype{a}{>{\columncolor{tablerowcolor}}r}
\definecolor{aicolor}{HTML}{018571}
\definecolor{occolor}{HTML}{ff7799}

\definecolor{aicolor}{HTML}{fc8d62}
\definecolor{occolor}{HTML}{253494}

\colorlet{tableheadcolor}{gray!25} 

\definecolor{neutralCol}{HTML}{dd1c77}
\definecolor{neutralGreen}{HTML}{31a354}
\definecolor{NewBlue}{HTML}{1879ba}
\definecolor{bleudefrance}{rgb}{0.19, 0.55, 0.91}  
\definecolor{AfTrColor}{HTML}{0868ac}  
\definecolor{BfTrColor}{HTML}{a8ddb5}  

\definecolor{AfCtColor}{HTML}{b10026}  
\definecolor{BfCtColor}{HTML}{fd8d3c}

\graphicspath{ {figures/} }

\newcommand{\para}[1]{\vspace{0.2em}\noindent\textbf{\textit{#1}~}}

\usepackage{xcolor}

\usepackage{newfloat}
\usepackage{listings}
\floatstyle{ruled}
\newfloat{listing}{tb}{lst}{}
\floatname{listing}{Listing}

\title{From Web(logs) to Web(AI): Questions, Platforms, and Methods\\ across Twenty Editions of ICWSM}

\author{
Koustuv Saha, Eshwar Chandrasekharan
}
\affiliations{
Siebel School of Computing and Data Science\\ University of Illinois Urbana-Champaign, Urbana, IL, USA\\
\{ksaha2, eshwar\}@illinois.edu}

\nocopyright
\begin{document}

\maketitle

\begin{abstract}
Over twenty editions, the ICWSM community has examined social life online as platforms, interactions, and research methods have changed. 
What can this body of research tell us at this critical juncture, as AI increasingly reshapes how people communicate online? 
We analyzed 2,139 indexed contributions from 2007 to 2026, distinguishing topics identified through nonnegative matrix factorization from problem framings captured through explicit textual cues. 
We find that platform mentions shift from \textit{blogs} toward \textit{Twitter} and, more recently, \textit{Reddit}. 
Online community research maintains a similar topic share (10.5\% to 10.0\%), but governance cues within it increase from 4.0\% to 34.3\%. 
Harm-related cues also increase after restricting abstracts to a fixed length. 
Our review also traces advances in sampling, measurement, and causal and experimental methods. 
We discuss how AI-mediated interactions complicate these questions and provide a reporting checklist to support research across changing platforms.


\end{abstract}

\section{Introduction}

Online social interaction has shifted across blogs, social networking sites, discussion forums, messaging services, and algorithmically curated feeds.
These changes complicate social media research because findings from one platform, population, or period may reflect the systems through which behavior was observed. AI adds another layer by suggesting, rewriting, generating, ranking, and moderating online communication~\citep{hancock2020,hohenstein2023}. Understanding what social media research has learned therefore requires tracking how its objects of study, questions, and methods have changed.

The International AAAI Conference on Web and Social Media (ICWSM) offers a useful record through which to examine these changes over two decades. 
Since its first edition in 2007, the conference has followed the development of online social life across different platforms and forms of interaction. Its research has examined self-expression, relationships, communities, support, public participation, and harmful behavior. It has also contributed methods for studying these phenomena, including evaluations of platform sampling~\citep{p14401}, adjustments for textual confounding~\citep{p19362}, randomized interventions~\citep{p19308}, and quasi-experimental designs~\citep{p31365}. ICWSM's twentieth edition therefore provides an opportunity to examine what has changed across this body of research, what has remained, and what these developments mean for studying increasingly AI-mediated interactions.

Prior studies show how the history of a research community can reveal changes in its intellectual interests, methodological commitments, and assumptions~\citep{hall2008,liu2014,wallace2017}. 
The FAccT retrospective, for example, examined how a growing research community defined its central concerns and forms of contribution~\citep{facct2022}. 
Within ICWSM, research on geographical representation has shown that diversity in platforms or research topics does not necessarily reflect diversity in the populations represented~\citep{weird2024}. 
These studies provide important accounts of what research communities study and whom their evidence represents. 
Accordingly, it is important and interesting to understand how ICWSM's topics, problem framings, platform contexts, and methodological approaches have evolved together, and what this history can inform about ongoing and future research.

In particular, for such a problem on studying a research community's publications, examining topics alone cannot provide this account. 
Researchers may study the same subject (e.g., social media) while asking different questions about it. 
For instance, research on online communities, may examine how participation develops, how members exchange support, how harmful behavior spreads, or how rules are enforced. 
In addition, methods shape what can be learned by determining whose behavior is observed, what a digital trace is taken to represent, and which comparisons support an explanatory or causal claim.
Therefore, towards disentangling various facets of ICWSM research, we ask the following research questions (RQs):

\begin{itemize}
    \item \textbf{RQ1:} How have ICWSM's research topics and platform contexts evolved across editions?
    \item \textbf{RQ2:} How have problem framings changed across and within research topics?
    \item \textbf{RQ3:} What methodological developments and recurring challenges are visible across changing platform contexts?
\end{itemize}

To answer these RQs, we analyze the titles and abstracts of 2,139 indexed contributions published across twenty editions of ICWSM from 2007 to 2026. We use nonnegative matrix factorization to estimate the distribution of research topics and a theory-informed codebook of textual cues to approximate problem framings. We examine changes across four fixed five-edition periods and conduct sensitivity analyses for abstract length, vocabulary change, and overlap between topic terms and framing cues. We complement this analysis with a targeted review of methodological contributions concerning sampling and representation, construct measurement, causal and experimental design, governance and disagreement, and research infrastructure.

Our results show that changes in research problems can occur even when the corresponding topic remains stable. Online community research accounts for a similar share of the topic distribution in the earliest and most recent periods, decreasing only from 10.5\% to 10.0\%. Within this research, however, governance-related cues increase from 4.0\% to 34.3\%. Harm-related framing also becomes more common, and this increase persists when abstracts are restricted to a fixed length. Platform contexts change more visibly: research moves from blogs toward Twitter and, more recently, toward Reddit, while work on language models grows in the latest editions. Our methodological review also traces advances in sampling, measurement, and causal and experimental methods, alongside recurring questions about representation, validity, governance, and data access.

This paper makes three contributions. First, we provide a documented corpus and longitudinal account of the topics and platform contexts represented across twenty editions of ICWSM. Second, we distinguish changes in what the community studies from changes in the problems it asks, showing how a relatively stable topic can acquire a different research framing over time. Third, we connect these changes to methodological developments and provide a reporting checklist for research on changing platforms and AI-mediated interactions. Together, these contributions show why the field's accumulated knowledge remains valuable, while also identifying the assumptions that researchers must reconsider as AI changes how online interactions are produced, encountered, and interpreted. We will release the corpus and analysis materials upon publication.
\section{Related Work}
\paragraph{Computational histories of research fields.}
\citet{hall2008} used topic models to follow ideas through computational linguistics. Venue retrospectives have also used co-word analysis at CHI \citep{liu2014} and traced technologies, methods, and values at CSCW \citep{wallace2017}. These studies show that a field's history depends on what evidence the retrospective follows.

\paragraph{Reflection within social computing.}
The FAccT retrospective combined publication analysis with a community questionnaire \citep{facct2022}. \citet{weird2024} examined 494 ICWSM papers from 2018--2022, retaining 420 for geographical analysis. They report that 37\% focused exclusively on Western populations, compared with 76\% at CHI and 84\% at FAccT. They also find that ICWSM studies still predominantly examine populations from countries that are more educated, industrialized, and rich than those studied at FAccT. We complement this population-focused analysis with a longer view of topics and problem framings.

\paragraph{Computational social science and its evidence.}
Foundational accounts describe the opportunities and institutional conditions for studying social behavior through digital traces \citep{lazer2009,lazer2020}. Others emphasize platform selection, data access, and the difficulty of treating observed users as a population \citep{ruths2014,p14517}. \citet{wallach2018} distinguishes social-scientific explanation from applying computational methods to social data. These concerns motivate our attention to the questions pursued within a topic.

\paragraph{Measurement, text, and causal inference.}
Topic-model fit does not establish that people find a component meaningful \citep{chang2009}, and automated text analysis requires substantive interpretation and validation \citep{grimmer2013}. Measurement theory further distinguishes an intended construct from its observable operationalization \citep{jacobs2021}. Dictionary estimates can also diverge from survey measures across populations \citep{jaidka2020estimating}. Text-based causal research faces additional problems when language represents confounders, treatments, or outcomes \citep{keith2020}. We draw on these distinctions when comparing framing cues with research designs. Our separation of topics and problems is also informed by problematization, which examines how scholars question assumptions within a body of work \citep{alvesson2011}. Framing cues approximate these questions; they cannot recover the full process through which a problem is formulated.

\section{Corpus and Methods}
\subsection{Scope and Retrieval}
We examine indexed main-conference contributions from 2007--2026, including full papers, shorter contributions, datasets, and demonstrations. We exclude workshops, tutorials, keynote summaries, and prefaces. 
We retrieved 2,233 records through Crossref's journal endpoint for ICWSM's online ISSN, 2334-0770, and checked volume structure against the AAAI archive.\footnote{\url{https://ojs.aaai.org/index.php/ICWSM/issue/archive}} 
We excluded 149 records in separate workshop issues, ten front-matter or program summaries, and seven workshop contributions in the 2017 main issue, leaving 2,067 contributions from 2008--2026.
We noted that the inaugural edition is absent from this Crossref collection. 
Therefore, we retrieved the official 2007 program\footnote{\url{https://www.icwsm.org/program.html}} and its 72 linked abstract pages spanning full papers, short papers, posters, and demonstrations. 
The resulting corpus contains \textbf{2,139 contributions}, all with a title and abstract. Conference year follows the numbered volume.
We remove HTML markup and normalize whitespace. 
The corpus is bounded by the retrieved sources; Appendix~\ref{app:corpus} reports annual counts.

\subsection{Platforms, Topics, and Problem Framings}
All main period comparisons use \textbf{2007--2011, 2012--2016, 2017--2021, and 2022--2026}, containing 386, 503, 488, and 762 contributions, respectively. 

\para{Platforms.} We use case-insensitive dictionaries to identify mentions of 20 platform families, along with a separate vocabulary for weblogs (Appendix~\ref{app:platforms}). The Twitter dictionary includes terms such as \textit{tweet} and \textit{retweet}. We also examined references to X from 2023--2026 and identified seven contributions that referred to the platform without mentioning Twitter.
In addition, we identify platform mentions from titles and abstracts. These mentions do not necessarily mean that a study analyzed data from that platform.

\para{Topics.} We represent each title and abstract using TF--IDF weights for unigrams and bigrams. We retain terms that appear in at least eight documents but in no more than 70\% of the corpus. After removing common scholarly and platform terms, the final vocabulary contains 3,837 features. We apply NMF with 15 components, NNDSVD-based initialization, a random seed of 42, and a maximum of 600 iterations. To examine the sensitivity of this choice, we compare models with 10, 12, and 18 components and rerun the 15-component model with three random initializations. The community component remains relatively stable across these runs, with its share in the most recent period ranging from 9.9\% to 10.7\%. Some other components are less stable, with minimum matched cosine similarities of 0.03 and 0.25 in two runs. We therefore treat the 15-component model as exploratory. We assign descriptive labels by examining each component's top terms and the abstracts with the highest weights. Appendix~\ref{app:topics} presents all topic labels and top terms.


For document-factor weights $W$ and term-factor weights $H$, we correct component scale before normalizing:
\[
q_{ik}=\frac{W_{ik}\sum_v H_{kv}}{\sum_\ell W_{i\ell}\sum_v H_{\ell v}}.
\]
Mean topic shares are $100N^{-1}\sum_iq_{ik}$ and sum to 100\%. They describe reconstructed TF--IDF mass. The median largest document weight is 0.45, motivating the use of mixtures.

\para{Problem framings.} We define eight overlapping problem framings: structure and diffusion, prediction and classification, measurement and validity, explanation and intervention, harm and information integrity, governance and participation, support and well-being, and resources and access. We use regular expressions to identify explicit cues for each framing in titles and abstracts. For each category, we report the percentage of contributions containing at least one corresponding cue. Appendix~\ref{app:codebook} provides the category definitions and example cues. We developed the codebook iteratively and examined 32 cue-positive abstracts to remove terms that were commonly used in unrelated contexts. Because this check included only abstracts that matched a cue, it does not measure recall or independently validate the categories. Therefore, we interpret these measures as the prevalence of framing cues, rather than the percentage of papers primarily concerned with each problem.


Within a topic, we calculate $100\sum_i q_{ik}b_{ip}/\sum_i q_{ik}$. This statistic weights papers by topical membership; its denominator differs from overall framing prevalence. Both axes derive from the same text and can share vocabulary. They are complementary text-derived measures.

\subsection{Sensitivity and Resampling}
Because abstract lengths may have changed over time, papers with longer abstracts could get more opportunities to match a framing cue. 
This could produce an apparent increase in framing prevalence even if the underlying research emphasis had not changed. To account for this possibility, we recalculate the framing rates using the title and first 100 words of each abstract, and then using only the first 100 abstract words. For abstracts shorter than 100 words, we use all available words without padding. We also calculate the number of harm-related cue matches per 100 words of title and abstract text.
For historical vocabulary, we add \textit{trolling}, \textit{flaming}, and \textit{vandalism} terms to the harm category in every edition; spam terms are already included. This check tests coverage under a common expanded vocabulary.

To assess shared vocabulary, we report overlap between each topic's twenty leading terms and all framing lexicons. For the community-governance comparison, we remove all seven governance-matching features from the document representation and fixed topic basis, then re-estimate and normalize document weights. This isolates direct lexical overlap without refitting the topic model.

We generate 2,000 resamples by sampling papers with replacement within each edition and recompute annual and pooled-period estimates with the topic model and codebook fixed. Reported 95\% intervals are the 2.5th and 97.5th percentiles. They describe sensitivity to paper composition in the retrieved corpus; coverage, coding, and model-selection uncertainty remain outside these intervals.
Appendix~\ref{app:boundaries} reports exploratory temporal segmentation. Additional checks vary topic count, initialization, text length, and title deduplication. Track composition remains a separate limitation.

\subsection{Methodological Comparison and Validation Scope}


Section~\ref{sec:methods_advances} organizes selected studies around sampling and representation,
construct measurement, causal and experimental design, governance and
disagreement, and research infrastructure. We selected 75 contributions for
methodological relevance and coverage across editions. They form an
illustrative bibliography and do not estimate conference-wide adoption or
influence. Appendix~\ref{app:validation} specifies independent validation procedures.
\section{Topics, Problem Framings, and Platforms}
\subsection{Changing Topics and Persistent Subjects}
\begin{figure*}[t]
\centering
\includegraphics[width=\textwidth]{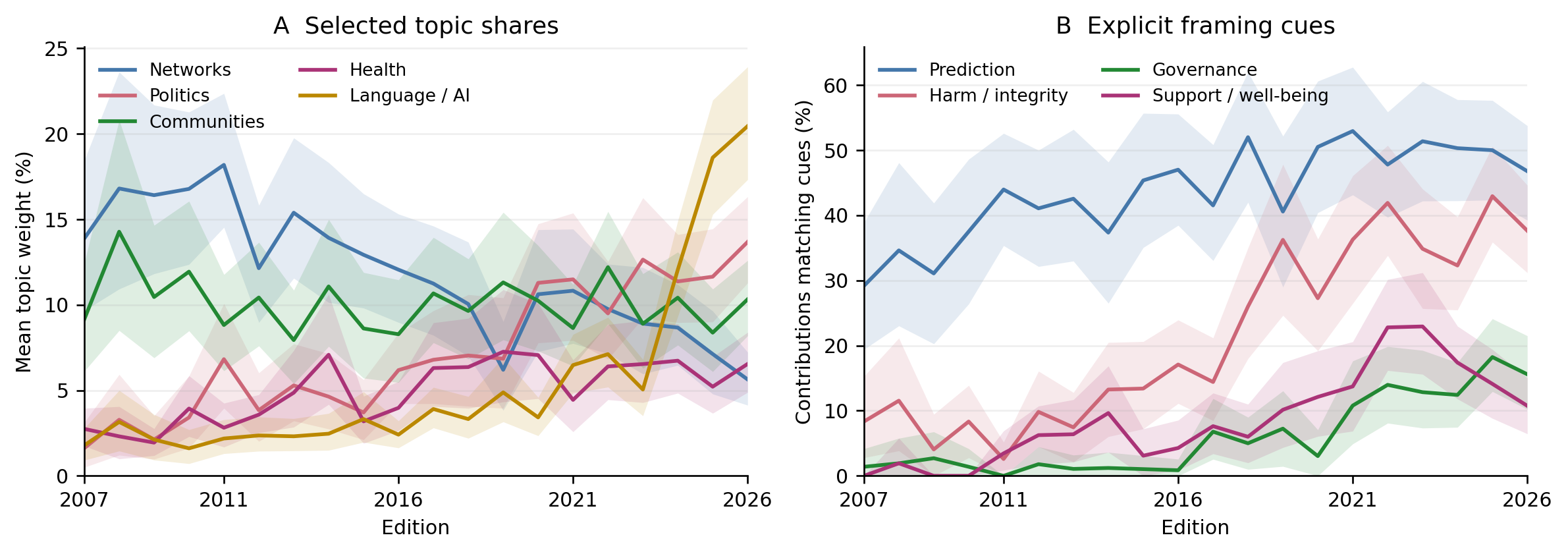}
\caption{Selected annual topics (A) and problem-framing cues (B). Topic shares describe mean NMF mixture weights; framing-cue rates describe overlapping textual matches. Bands are 95\% within-edition resampling intervals. They reflect composition sensitivity under fixed measures; they do not assess semantic validity.}
\label{fig:keytrends}
\end{figure*}

Figure~\ref{fig:keytrends} presents selected annual series for RQ1 and RQ2; Appendix~\ref{app:topics} gives the complete distributions. Across the first and last five-edition periods, the networks/diffusion/recommendation component decreases from 16.6\% to 7.8\% of mean topic weight. Search, question answering, and tagging decrease from 11.0\% to 3.2\%, and sentiment/opinion mining from 10.5\% to 4.1\%. Politics/polarization increases from 3.8\% to 11.8\%, and health/social support from 2.8\% to 6.3\%. These are changes in relative composition; a falling share can coexist with continued publication on a subject.

The language-modeling component reaches 20.5\% in 2026. Small early-period weights reflect shared language and text terms and are not evidence of early generative-AI research. A narrower lexical measure finds explicit LLM/generative-AI terms in 23/161 contributions in 2024 (14.3\%), 41/170 in 2025 (24.1\%), and 61/186 in 2026 (32.8\%). These counts measure visibility in the proceedings, not the prevalence of AI-generated content on social media.

\subsection{Different Problems within Community Research}
The community component's mean share changes little: 10.5\% [8.9, 12.2] in 2007--2011 and 10.0\% [8.8, 11.2] in 2022--2026, an endpoint difference of $-0.5$ percentage points [$-2.6$, 1.5]. Brackets throughout this section denote 95\% resampling intervals.

\begin{figure*}[t]
\centering
\includegraphics[width=\textwidth]{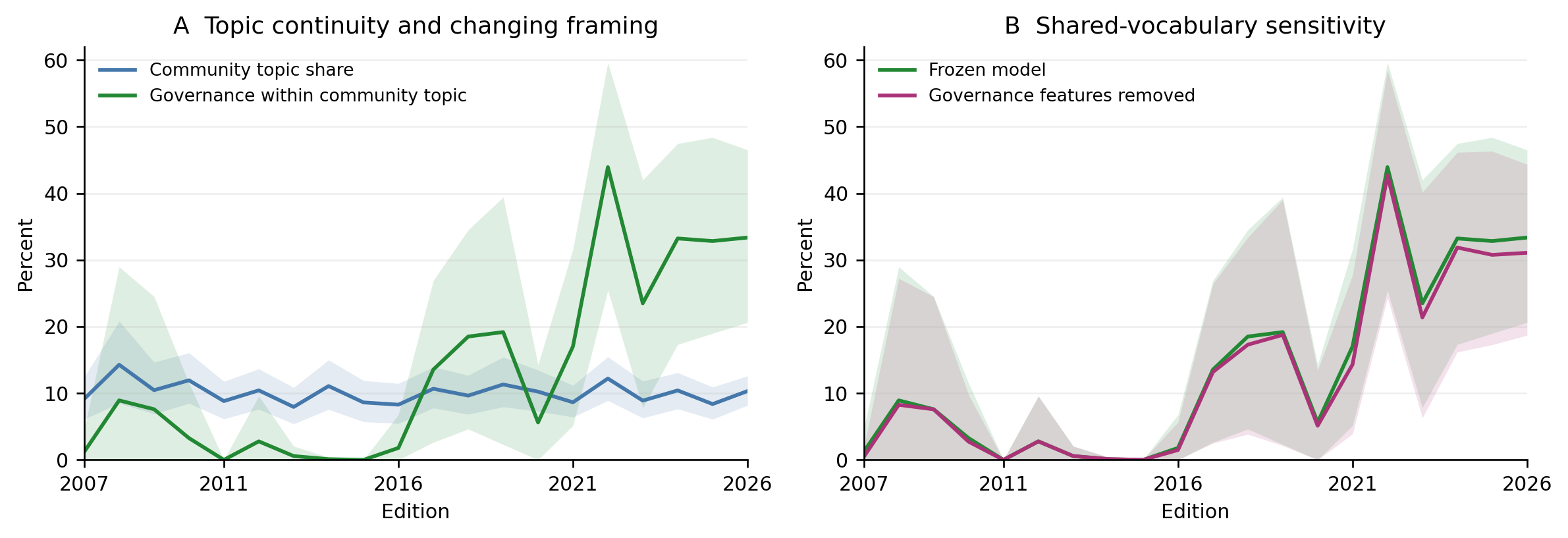}
\caption{Community research and governance framing. Panel A compares the community topic's share of all topic weight with the fraction of community-topic weight attached to papers containing governance cues; these series have different denominators. Panel B repeats the conditional rate after removing governance-matching features before projection onto the fixed topic basis. Bands describe composition sensitivity under the corresponding fixed representation.}
\label{fig:community}
\end{figure*}

Governance cues within community-topic weight increase from 4.0\% [0.2, 8.8] to 34.3\% [27.0, 41.4], a difference of 30.3 points [21.8, 38.4]. Figure~\ref{fig:community} shows the annual series. It is uneven, but the endpoint contrast is large.

Direct vocabulary overlap leaves the contrast largely intact. \emph{Moderation} is one of the community component's twenty leading terms and also matches the governance lexicon. Removing all seven governance-matching features changes the endpoint comparison to 3.6\% [0.1, 8.2] versus 32.5\% [25.3, 39.6], a difference of 28.9 points [20.6, 36.8]. Correlated vocabulary still links the two text-derived measures.

The same pattern survives stricter text checks. Restricting governance cues to the title and first 100 abstract words gives 2.5\% [0.0, 6.6] versus 25.8\% [18.7, 32.6], a difference of 23.2 points [15.1, 31.1]. Excluding papers whose only governance cues are \textit{moderate}, \textit{moderated}, \textit{moderates}, or \textit{moderating}, which often mark statistical or adjectival uses, gives 2.5\% [0.0, 6.6] versus 33.3\% [26.0, 40.4]. Refitting topics on titles and the first 100 abstract words leaves community share at 10.3\% and 9.7\% in the endpoint periods. Platform composition does not account for the contrast either. Standardizing each period to the 2007--2011 platform composition gives 4.0\% versus 29.7\%, a difference of 25.7 points [15.2, 35.6], and the rate holds at 4.3\% versus 30.1\% among contributions mentioning neither Twitter nor Reddit (Appendix~\ref{app:sensitivity}).

\subsection{Harm Framing and Measurement Sensitivity}
Across all contributions, harm/information-integrity cues increase from 6.2\% [3.9, 8.5] to 38.1\% [34.8, 41.2]. Governance cues increase from 1.3\% to 14.8\%, support/well-being cues from 1.3\% to 16.8\%, and prediction/classification cues from 36.3\% to 49.1\%. Table~\ref{tab:periods} uses the same four periods for all selected series.

\begin{table*}[t]
\small
\centering
\begin{tabular}{@{}lrrrr@{}}
\textbf{Measure} & \textbf{2007--2011} & \textbf{2012--2016} & \textbf{2017--2021} & \textbf{2022--2026} \\
\toprule
\rowcolmedium Contributions ($N$) & 386 & 503 & 488 & 762 \\
\hdashline
Networks / diffusion & 16.6 & 13.2 & 10.1 & 7.8 \\
Politics / polarization & 3.8 & 4.8 & 8.7 & 11.8 \\
Communities / participation & 10.5 & 9.2 & 10.0 & 10.0 \\
Health / social support & 2.8 & 4.4 & 6.2 & 6.3 \\
Prediction / classification cues & 36.3 & 42.9 & 47.7 & 49.1 \\
Measurement / validity cues & 8.5 & 15.1 & 15.0 & 18.1 \\
Explanation / intervention cues & 2.1 & 1.4 & 7.6 & 10.5 \\
Harm / integrity cues & 6.2 & 12.3 & 27.0 & 38.1 \\
Governance cues & 1.3 & 1.2 & 6.6 & 14.8 \\
Support / well-being cues & 1.3 & 5.8 & 9.8 & 16.8 \\
Governance within community-topic weight & 4.0 & 1.2 & 14.4 & 34.3 \\
\bottomrule
\end{tabular}
\caption{Selected measures across the same four five-edition periods. All values except $N$ are percentages. The first four rows of measures are mean topic weights; the next six are overlapping cue prevalences. The last row is conditional on community-topic weight. Headline resampling intervals are reported in the text.}
\label{tab:periods}
\end{table*}

\textbf{Abstract length.} Median abstract length increases from 113 words in 2007 to 190 in 2026. With titles plus the first 100 abstract words, harm-cue prevalence changes from 5.7\% [3.4, 8.0] to 31.9\% [28.7, 35.0]. With only the first 100 abstract words, it changes from 5.7\% [3.4, 8.0] to 31.0\% [27.8, 34.1]. The latter difference is 25.3 points [21.5, 29.2], smaller than the full-text difference. Mean harm-cue density also increases, from 0.10 to 0.78 matches per 100 title/abstract words.

\textbf{Historical vocabulary.} Adding trolling, flaming, and vandalism terms leaves the early-period harm estimate at 6.2\% and changes the last-period estimate to 38.3\%. Historical coverage remains uneven: 42.5\% of early contributions match none of the eight categories, compared with 14.0\% in the last period. Period-specific coding is needed to separate changing terminology from changes in framing. Appendix~\ref{app:sensitivity} reports the alternative measures for every period.

\subsection{Platform Context}
\begin{figure*}[t]
\centering
\includegraphics[width=\textwidth]{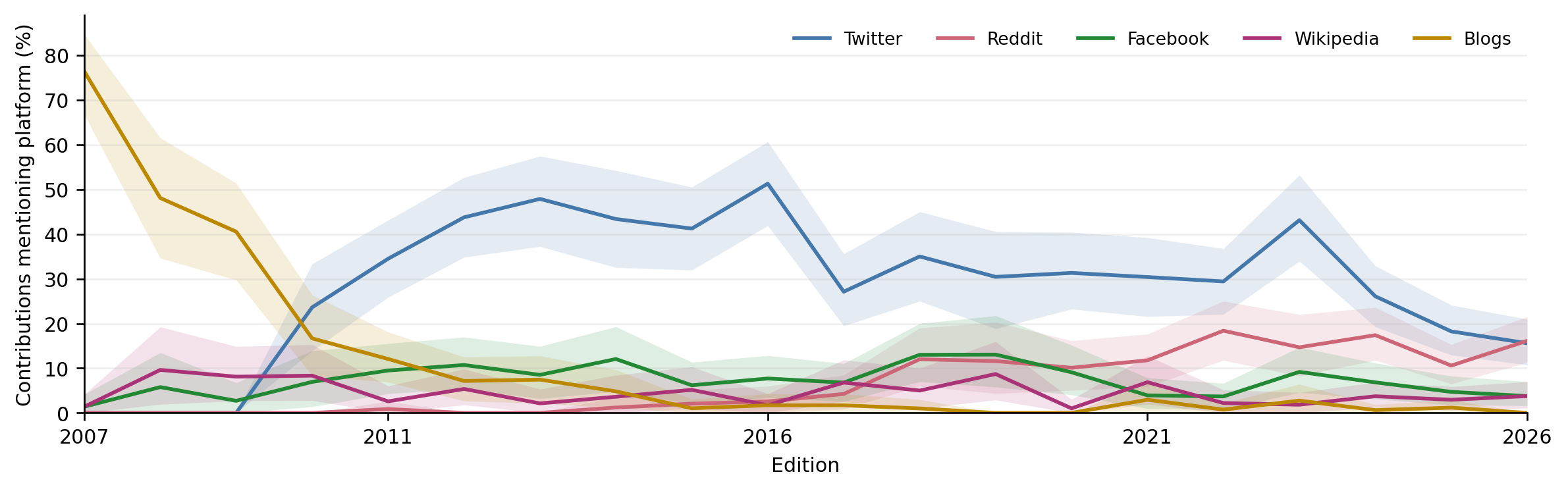}
\caption{Platform mentions in titles and abstracts for selected long-running contexts, with 95\% composition-resampling bands. Counts can overlap across platforms. The full dictionary contains twenty named families plus weblog terms (Appendix~\ref{app:platforms}).}
\label{fig:platforms}
\end{figure*}

Figure~\ref{fig:platforms} addresses the platform dimension of RQ1. Blog mentions decrease from 35.2\% in 2007--2011 to 0.9\% in 2022--2026. Twitter-associated mentions reach a period peak of 45.7\% in 2012--2016 and decline to 24.8\% in 2022--2026. Reddit increases from 1.2\% in 2012--2016 to 15.4\% in 2022--2026. In 2026, Twitter/X and Reddit appear in 15.6\% and 16.1\% of contributions, respectively. These counts describe changing publication contexts; they do not measure platform replacement or equivalent data access.

Decentralized platforms have a small, uneven presence in this corpus. Mastodon is mentioned in 0, 3, 0, and 3 papers across 2023--2026; Bluesky in 0, 0, 2, and 3. Publication timing and title/abstract matching limit what these small counts can show.

Annual profiles do not support sharp phase boundaries: no boundary in the baseline five-segment model appears in more than 65\% of 200 resamples. We retain fixed comparison periods; Appendix~\ref{app:boundaries} reports the segmentation diagnostics.

\section{Methodological Developments and Recurring Questions}
\label{sec:methods_advances}
Our targeted review examines how ICWSM researchers have addressed recurring methodological questions as platforms and available data have changed. We focus on five areas: sampling and representation, construct measurement, causal and experimental design, governance and disagreement, and research infrastructure. The studies illustrate developments and continuing challenges within each area; they do not measure how common these approaches are across the conference.

\subsection{Sampling and Representation}
Large datasets do not remove the need to specify whose behavior is observed. Early ICWSM work documented uneven Twitter demographics \citep{p14168}. Sampling later became an object of direct comparison: \citet{p14401} compared Twitter's sampled stream with Firehose data, and \citet{p7337} showed that fidelity depends on the scale and unit measured. 
Platform partnerships expose per-user activity logs, including cyclic routines that public interfaces do not capture~\cite{chowdhury2021ceam}.
Some questions also require observations collected outside platform interfaces.
Browsing histories from tens of thousands of users showed that visits to politically aligned news sources last substantially longer, a pattern that the hyperlink structure of the Web alone does not explain~\cite{garimella2021political}.
Community selection raises a related problem. Demographic reweighting, for example, need not improve population prediction \citep{p19287}, while multilingual work broadens the contexts examined without automatically making them representative \citep{p19271}. 

\subsection{Construct Measurement}
A trace can support more than one interpretation. A link may reflect attention, affiliation, or a relationship; a post may express distress without measuring clinical severity. In 2007, \citet{icwsm2007_paper19} combined network analysis with a survey of blogging relationships, and \citet{icwsm2007_paper30} compared personality impressions with self- and acquaintance reports. Both used evidence beyond the trace to assess its meaning.

Later studies revisited familiar proxies. \citet{p14033} distinguished follower counts from retweet- and mention-based influence. 
Conversational resilience has likewise been operationalized from what happens after a norm violation, including whether discussion continues and whether later outcomes are adverse or prosocial~\citep{lambert2022conversational}.
Further, disaggregating incivility into vulgarity, name-calling, aspersion, and stereotypes shows that categories collapsed by a single toxicity score relate differently to subsequent participation~\cite{gao2024crisis}.
CREDBANK paired streaming tweets with event-level crowd credibility judgments, so that credibility could be measured against human ratings~\cite{mitra2015credbank}. 
Likewise, collective attention has been operationalized from the descriptors people use when referring to an event, which shift as shared knowledge accumulates~\cite{stewart2020characterizing}.
\citet{p14661} examined the meanings assigned to retweets through a user survey and a metareview. 
Electoral prediction also exposed problems of transfer across settings \citep{p14189,p14773}. These studies make validation specific to a construct, population, and use.

Lexicon-based methods made many social and psychological constructs measurable at scale. Linguistic Inquiry and Word Count (LIWC) maps words to psychological and social categories~\cite{tausczik2010psychological}, and early ICWSM work applied it to English and Spanish depression forums~\cite{p18623}, and understanding activist movements~\cite{de2016social}. 
VADER adapted sentiment scoring to social media conventions (e.g., emoticons, slang, and capitalization)~\cite{hutto2014vader}, and CrisisLex used curated terms to collect crisis communication~\cite{p14538}. These resources offered transparency, low cost, and comparability across studies. Their validity also depended on context. 
Lexicon-based hate-speech detection conflated offensive language with hate speech~\cite{davidson2017automated}, and dictionary-based estimates of regional well-being diverged sharply from survey measures because word use varies across populations~\cite{jaidka2020beyond}. 
Coordinated political campaigns also circulate lexically mutated variants of the same message, which fixed keyword lists match unevenly~\cite{phadke2024characterizing}.
A comparison on Chinese text found local moral-foundation lexicons insufficient relative to multilingual and large language models, while still calling for human validation of model outputs~\cite{cheng2026beyond}. The validation question thus extends from word lists to prompts, training data, and model versions. Our framing indicators share the strengths and limits of this tradition: they are transparent and reproducible, and their meaning requires validation against human judgment.




\subsection{Causal and Experimental Design}
\label{sec:causal}
Social media histories provide pretreatment language and behavior, but causal claims require a defensible comparison. Table~\ref{tab:designs} contrasts randomized interventions, quasi-random assignment, difference-in-differences, estimator evaluation, and matching-based designs represented in ICWSM.

\begin{table*}[t]
\centering
\small
\begin{tabular}{@{}p{.14\textwidth}p{.21\textwidth}p{.28\textwidth}p{.28\textwidth}@{}}
\textbf{Design} & \textbf{ICWSM Example} & \textbf{Requires} & \textbf{Leaves open} \\
\toprule

Randomized intervention &
Reconsideration prompts on Twitter \citep{p19308} &
Random assignment supports a comparison of eligible users assigned to receive the prompt. &
Classifier-defined eligibility and outcomes; transfer beyond English replies and the study setting. \\
\hdashline

Quasi-random assignment &
Civil communication on Roblox \citep{p31365} &
Residual stochasticity in matchmaking assignment acts as a quasi-randomization mechanism for the analyzed server comparisons. &
Other server properties may co-vary with civility; validity and transfer of the local comparison. \\
\hdashline

Matching and difference-in-differences &
Fringe-platform participation \citep{p22184} &
Comparable untreated trends after matching; no differential concurrent changes explaining the contrast. &
Selective migration, cross-platform linkage, and changing composition or exposure. \\
\hdashline

Regression discontinuity / interrupted time series &
Conversation threading after a platform interface change
\citep{aragon2017thread} &
A defined intervention point supports comparison of outcomes before and after the platform change. &
Concurrent changes and time-varying factors may also explain the observed discontinuity. \\
\hdashline

Text-adjustment evaluation &
Known-effect benchmark tasks~\cite{p19362} &
Effects are known under the benchmark's construction, enabling comparisons of estimators. &
Performance under the benchmark need not transfer to an unknown real-world assignment process. \\
\hdashline

Stratified propensity-score matching &
Support language~\citep{p14891}, college alcohol mentions~\citep{p15012},
psychiatric medication use~\citep{saha2019social}, and algorithmic rank
movement~\citep{chan2026examining} &
Conditional exchangeability, overlap, consistent exposure definitions,
and appropriate timing. &
Unmeasured circumstances and linguistic proxies; balance alone cannot
establish exchangeability. \\
\hdashline



Balanced risk-set matching &
Cross-community fringe interactions~\citep{russo2024stranger} &
Comparable treated and untreated users can be identified within aligned risk sets using only pre-treatment information. &
Unmeasured confounding may remain; estimates depend on how text,
context, and eligible comparison users are represented. \\
\hdashline



Outcome-oriented case--control comparison & Psychosocial outcomes on TalkLife \citep{p7326} & Comparisons are conditional on subsequent outcomes and observed behaviors in groups. & Outcome selection and residual confounding; does not itself identify an assigned intervention's effect. \\

Randomization under interference &
Cascade-based design \citep{p31322} &
The network and diffusion model adequately describe relevant interference in the evaluated setting. &
Misspecified or changing networks; simulation performance does not establish a field effect. \\

\bottomrule
\end{tabular}
\caption{Causal-inference designs in ICWSM research, with the conditions each design requires and the questions it leaves open. Examples illustrate each design's use.}
\label{tab:designs}
\end{table*}

The examples make exposure definitions and comparison groups explicit. 
These studies have rationalized the need for quasi-experimental methods when conducting a randomized controlled trial is infeasible or impractical~\cite{p15016}.
Engagement with a counseling post, for example, differs from reading it or receiving counseling~\cite{p15016}. 
Matching addresses measured differences, while unmeasured confounding can remain~\cite{p14891,p15012}. 
Temporal comparisons require an appropriate counterfactual trajectory, and network settings raise interference questions~\cite{saha2025mental}. \citet{p19362} also show why text-adjustment estimators need evaluation against known-effect tasks. 
Suggested replies and automated moderation create the same design problem: studies need explicit exposures and credible comparisons, with spillovers addressed when interactions cross users.
Audits of curation systems can also treat the system's ranking decision as the exposure: matched feed snapshots estimate how upward rank movement affects subsequent engagement \citep{chan2026examining}.

\subsection{Governance and Disagreement}
Questions about judgment and authority appear throughout ICWSM's history. Automatic comment moderation appeared in the inaugural proceedings \citep{icwsm2007_paper59}, and Wikipedia work examined governance through promotion processes \citep{p14013}. Later research connected community rules to perceived governance~\citep{p15033}. 
Governance also operates through interventions short of removal: soft moderation attached warning labels to a large share of election-related posts, with uneven application across accounts~\cite{zannettou2021won}. 
Across thousands of communities, the rules a community writes relate systematically to how members perceive its governance~\cite{leibmann2025reddit}.
These examples place classification decisions inside the institutions and norms that give them meaning.
Community feedback offers another signal of norms: highly upvoted content has been used as a proxy for what communities value and encourage~\cite{goyal2026uncovering}. Algorithmic exposure can also create moderation pressure: posts reaching Reddit's r/popular drew substantially more comments, newcomers, and removed comments than other posts from the same communities
\citep{chan2024understanding}.

LLM-based annotation and moderation extend those questions to model-generated judgments. Studies document LLM annotation bias \citep{p35879}, moderator disagreement \citep{p42625}, and how LLMs apply Wikipedia neutrality norms \citep{p42630}. A model's agreement with a reference label depends on whose judgment the label represents. Evaluations should document whose labels define the reference and where those decisions are used.

\subsection{Research Infrastructure}
Tools and datasets shape which studies can be conducted and repeated. CrisisLex supported research on crisis communication~\citep{p14538}; the Pushshift Reddit dataset expanded access to community activity~\citep{p7347}; Media Cloud and the Methods Hub support news archives and executable methods~\citep{p18127,p42804}. Infrastructure also sets conditions on use. Researcher access under the Digital Services Act makes those constraints part of the research question \citep{p42671}, while work on generative AI in crowdwork raises provenance concerns for apparently human-produced inputs~\citep{p31452}. In AI-mediated settings, provenance includes model versions and whether content was composed, edited, selected, or generated.

\section{Discussion}

Across twenty editions, ICWSM has followed online social life through substantial changes in platforms, technologies, and forms of interaction. Our findings show that these changes occur at different levels. Platform contexts can change visibly, as seen in the movement from blogs to Twitter and, more recently, toward Reddit and language-model research. Changes within research topics can be less visible.




\paragraph{Stable topics can contain changing research questions.}
Online community research accounts for a similar share of the topic distribution in the earliest and most recent periods, while governance has become a much more common framing within this research. 
This illustrates how a stable topic can
contain changing questions, outcomes, and stakeholders.
Understanding the development of a field therefore requires examining both what it studies and the questions it asks about those subjects.

Our phase analysis also does not support a clean division of ICWSM's history into distinct periods. Research interests often overlap, with some emerging while others continue or gradually decline. 
Publication records can show when particular questions received greater attention, but they cannot explain why these changes occurred. Future work could examine conference archives and calls for papers, changes in platform and data access, and interviews with members of the ICWSM community to better understand these shifts.

\paragraph{Methodological advances retain important assumptions.}
ICWSM research has made substantial methodological advances in sampling, measurement, causal inference, experimentation, governance, and research infrastructure. At the same time, each approach depends on assumptions that shape what can be concluded. Sampling determines which people and behaviors become visible. Digital traces require evidence that they represent the construct being studied. Experiments and matched comparisons require clearly defined treatments, outcomes, and comparison groups. Governance research depends on whose judgments define harmful or acceptable behavior. Data access determines which interactions can be observed and which remain outside the analysis.

These questions recur because new methods address particular sources of uncertainty without resolving every limitation. For example, comparing users within the same college subreddit can account for some regional, seasonal, academic-calendar, and local conditions. Participation in the same online community, however, cannot capture every difference in people's offline experiences. Making these assumptions explicit helps readers understand what a design addresses and what remains uncertain.

\paragraph{Making methodological decisions easier to evaluate.}
The recurring questions identified in our review motivate the reporting checklist in Table~\ref{tab:checklist}. The checklist asks researchers to describe the population their data represent, the quantity they seek to estimate, the meaning assigned to digital traces, the construction of comparison groups, and the sensitivity of findings to alternative decisions. It also asks researchers to document collection conditions, platform access, and system versions. These details allow future readers to assess whether a finding is likely to hold when the platform or data-generating process changes. The checklist is intended to make methodological assumptions visible; it does not establish that those assumptions are valid.

\begin{table*}[t]
\small
\centering
\begin{tabular}{@{}p{.20\textwidth}p{.73\textwidth}@{}}
\textbf{Reporting item} & \textbf{Information to provide} \\
\toprule
Target population & Who or what is the study about? Describe observed users, communities, languages, dates, and exclusions; distinguish the sample from the target population. \\
\rowcollight Question and estimand & State the descriptive quantity or causal contrast, unit of analysis, exposure, outcome, and time horizon. Specify whose effect is estimated. \\
Constructs and measures & Explain why each trace or label measures the intended construct. Report validation, disagreements, missingness, and relevant subgroup or period differences. \\
\rowcollight Collection and provenance & Record interfaces, sampling rules, access restrictions, deletion or attrition, and data and model versions. Document content-production processes where known. \\
Assignment and timing & Describe how exposure arises, the comparison group, and the ordering of covariates, treatment, and outcomes. Identify concurrent events and possible spillovers. \\
\rowcollight Adjustment and overlap & Justify pretreatment covariates, including text features. Report balance, overlap, exclusions, and any change in the target population after adjustment. \\
Sensitivity and uncertainty & State assumptions and plausible failures. Use appropriate sensitivity analyses, alternative measures, placebo or negative-control comparisons, and uncertainty estimates. \\
\rowcollight Interpretation and reuse & Separate prediction, association, and causal interpretation. State limits of transfer and provide reusable code, definitions, and access information when possible. \\
\bottomrule
\end{tabular}
\caption{A reporting checklist for studies using social media observations. Items should be applied to the study's question and design, with non-applicable items explained. The checklist is a reporting aid, not a quality score or a substitute for validation.}
\label{tab:checklist}
\end{table*}

\paragraph{Studying interaction when AI participates in communication.}
AI-mediated interaction makes the relationship between behavior and digital traces more complicated. AI systems can modify, augment, or generate messages on a person's behalf~\citep{hancock2020}. A message may be written entirely by a person, selected from suggested responses, edited from generated text, or produced with little direct input. The final text can therefore reflect the person's intentions, the model's output, interface defaults, and platform rules. Examining the text alone may not reveal how these different influences shaped it.

This distinction matters when researchers infer attitudes, distress, social support, or interpersonal behavior from language. Experiments with suggested replies show that AI assistance can change both the language people use and how their interaction partners evaluate them~\citep{hohenstein2023}. More positive language does not necessarily indicate that the person feels more positive, and a supportive-sounding message does not necessarily mean that its recipient experiences greater support. Future research should therefore distinguish among properties of the language, the intentions and experiences of the sender, the perceptions of the recipient, and the outcomes of the interaction. Our counts of language-model research describe growing scholarly attention to these systems; they do not estimate how much online communication is currently mediated by AI.

AI also changes how exposure should be defined. A person may encounter generated content, suggested replies, ranking decisions, or moderation actions, each of which can change across users and system versions. Studies of these interactions should document which content was generated, displayed, selected, edited, and received. They should also record the model and interface version, the interaction context, and whether participants recognized the involvement of AI. With appropriate consent and privacy protections, process records, participant accounts, and interface experiments can provide evidence that the final text alone cannot recover.

\paragraph{Preserving knowledge across changing systems.}
Changes in platforms, access policies, model versions, and content-production processes affect both online behavior and researchers' ability to observe it. A finding may change because people's behavior has changed, because the platform now structures interaction differently, or because researchers received access to a different part of the system. Recording data-access conditions, validation evidence, disagreement, system versions, and analytical decisions helps later researchers distinguish among these possibilities.

Social media increasingly consists of interactions among people, AI systems, and platforms. This shift makes ICWSM's longstanding questions about representation, measurement, causality, governance, and access increasingly relevant. Carrying this knowledge forward will require updating how these concepts are measured while preserving enough information to compare findings across changing systems.
\section{Limitations and Future Directions}
Our topic labels and framing cues approximate the constructs of interest. Components can combine domains, tasks, and writing styles; a cue can describe background. Length, vocabulary, and strict-cue checks address specific artifacts but cannot establish precision or recall. Independent judgments are needed for positive and negative cases. Appendix~\ref{app:validation} specifies a stratified coding and intrusion-test protocol; those results are not incorporated in the present estimates.

The corpus combines publication formats whose representation may change over time. Including datasets and demonstrations captures research infrastructure, but their abstracts may emphasize different problems from full papers. A full-paper-only analysis should use verified proceedings section labels and compare like formats within the same editions. Reconciliation of every issue and track would also strengthen the coverage audit.

The methodological examples come from an exploratory bibliography and cannot estimate the prevalence or influence of a design. A structured full-text review that samples beyond cue-positive abstracts and documents disconfirming cases would strengthen the synthesis.

Publication year is an imperfect proxy for when a study was conceived or its data were collected. Topic changes and segmentation boundaries do not identify why research interests changed. Platform contexts also changed over this period. Standardizing periods to a fixed platform composition, and restricting to contributions mentioning neither Twitter nor Reddit, both retain about five sixths of the community-governance contrast (Appendix D), but title and abstract mentions are a coarse proxy for the platform a study actually analyzed. Our resampling intervals hold measurements fixed and therefore exclude model and interpretation uncertainty. Future work could combine validated coding, alternative representations, citation analysis, and accounts from researchers. For AI-mediated interaction, publication trends also need direct evidence about how people and communities use the systems being studied.

\section{Conclusion}
Across 2,139 contributions and twenty ICWSM editions, we find platform turnover alongside recurring research questions. Community research stays near one tenth of topic weight, while governance cues within it rise from 4.0\% to 34.3\%. The contrast remains after removing shared governance vocabulary, shortening abstracts, and tightening the governance cue. Selected studies across the conference repeatedly confront problems of representation, measurement, causal comparison, governance, and data access. Platforms change; the questions recur. AI systems now participate in producing, ranking, and moderating online interaction, making the provenance of both content and exposure part of the evidence. We will release the corpus and analysis materials upon publication; the reporting checklist provides a common record for future comparisons.

\section*{Ethical Considerations}
Our study uses public scholarly publications and Crossref metadata. It involves no new recruitment or analysis of individuals' social media posts. We retain author names for bibliographic attribution and do not infer sensitive attributes or rank researchers. Publication counts and framing cues could be misused as indicators of research quality or community concern. We reduce this risk through explicit measurement definitions, counterexamples, and sensitivity analyses. Upon publication, we will release bibliographic metadata, machine-readable derived measurements, analysis code, dictionaries, and documentation with persistent identifiers, source links, and a data card describing provenance and reuse constraints. Crossref bibliographic metadata is reusable without restriction, while abstracts remain subject to publisher or author copyright.\footnote{\url{https://www.crossref.org/documentation/retrieve-metadata/}} We will not redistribute source papers or bulk abstracts.

\section*{Acknowledgments}
OpenAI's ChatGPT and Codex, and Anthropic's Claude, were used to assist with literature retrieval and synthesis, corpus assembly, AI-assisted inspection of topic labels and lexical cues, source checking, and manuscript drafting and revision. Quantitative results were computed with the Python analysis scripts. The authors directed the research design, reviewed the generated material, and take responsibility for the final content.

\fontsize{9pt}{8pt} {\selectfont
\bibliography{0paperICWSM}}

\appendix
\onecolumn
\setcounter{figure}{0}
\setcounter{table}{0}
\renewcommand{\thefigure}{A\arabic{figure}}
\renewcommand{\thetable}{A\arabic{table}}
\section{Corpus and Platform Dictionaries}
\label{app:corpus}
\begin{figure}[ht]
\centering
\includegraphics[width=.75\textwidth]{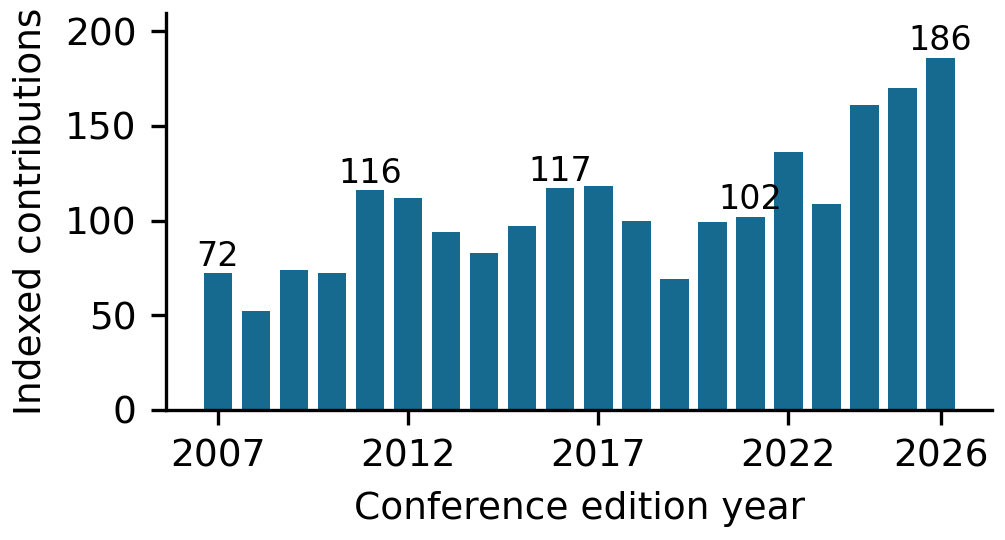}
\caption{Annual contributions in the retrieved corpus.}
\label{fig:corpus}
\end{figure}

Figure~\ref{fig:corpus} reports indexed contributions by edition. These are publication counts, not submission or acceptance counts. The main analysis includes all retained main-conference formats.
\subsection{Platform families}
\label{app:platforms}
The twenty named platform families are Twitter/X, Reddit, Facebook, Wikipedia, YouTube, Instagram, TikTok/Douyin, Weibo, WeChat, WhatsApp, Telegram, Gab, 4chan, Parler, Mastodon, Bluesky, Flickr, Digg, MySpace, and StackOverflow. A separate weblog indicator matches blogs, weblogs, blogging, bloggers, and blogosphere. Twitter includes tweets and retweets; Reddit includes subreddit and redditor variants; Wikipedia includes Wikipedian variants. StackOverflow allows a space between the two words. Seven X-only cases, identified through AI-assisted inspection, supplement the Twitter vocabulary. All other named families use the explicit name, with case-insensitive word boundaries. The regular expressions used in the analysis will be released upon publication and provide the exact specification. Rates count mentions, can overlap, and do not establish data use.

\section{Topic Components and Complete Distributions}
\label{app:topics}
\begin{figure}[p]
\centering
\includegraphics[width=\textwidth]{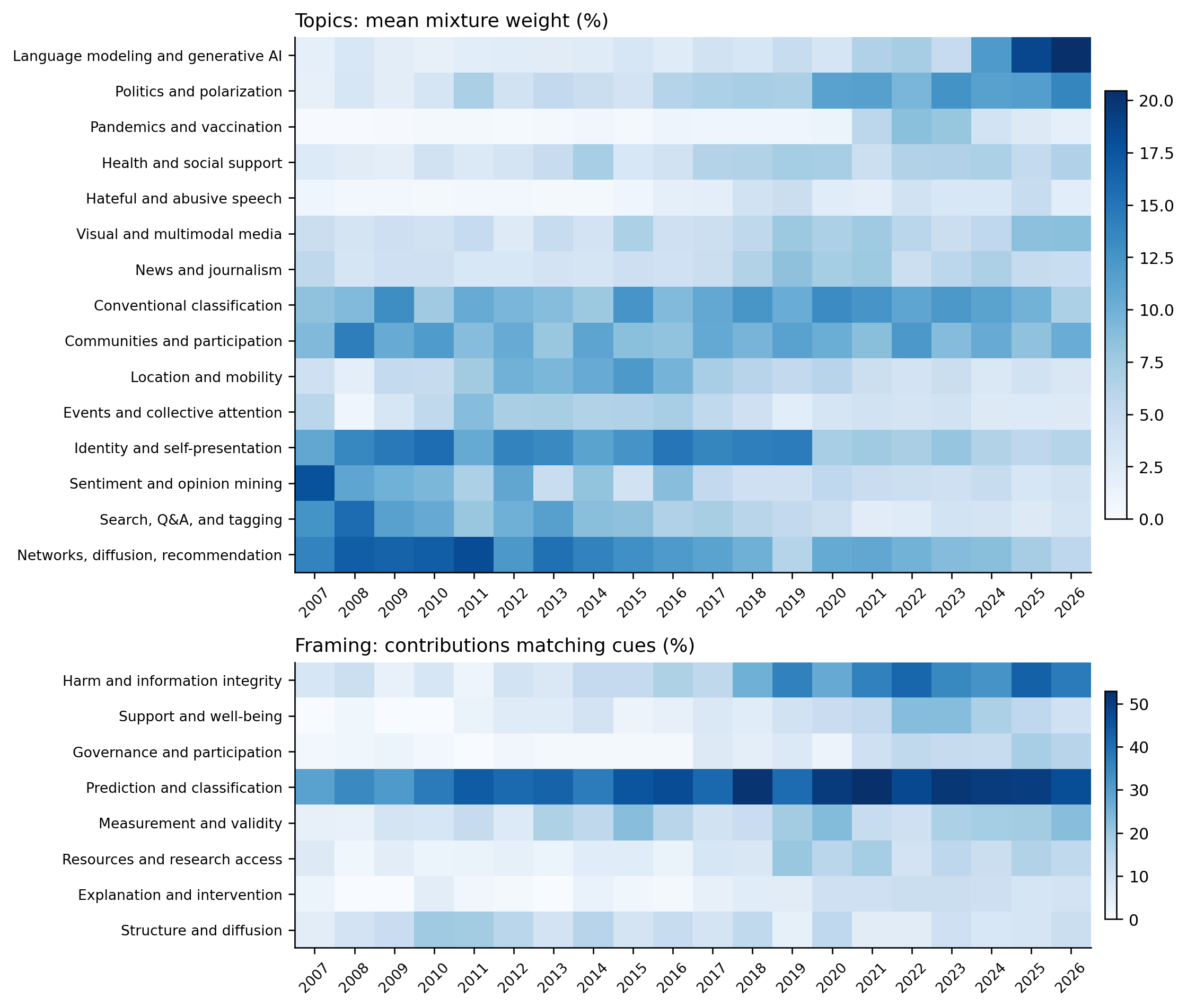}
\caption{Complete annual topic shares and framing-cue prevalences, ordered by endpoint-period change. Topic shares sum to 100\%; framing categories overlap. No historical phase boundaries are imposed.}
\label{fig:topicsproblems}
\end{figure}

Topic labels describe one exploratory 15-component NMF representation. Table~\ref{tab:topicterms} lists each component's twenty highest-weight terms in order. Figure~\ref{fig:topicsproblems} orders rows by the difference between the last and first fixed periods, from the largest increase to the largest decrease, separately for topics and framings. Component weights and cue prevalences have different denominators.
\begingroup
\small
\begin{longtable}{@{}p{.05\textwidth}p{.22\textwidth}p{.65\textwidth}@{}}
\caption{NMF components and top terms, ordered by component identifier.}\label{tab:topicterms}\\
\toprule
ID & Interpretive label & Twenty highest-weight terms \\
\midrule
\endfirsthead
\toprule
ID & Interpretive label & Twenty highest-weight terms \\
\midrule
\endhead
\bottomrule
\endfoot
T00 & Networks, diffusion, recommendation & diffusion, influence, link, graph, structure, recommendation, nodes, problem, prediction, links, cascades, node, systems, graphs, real, items, spread, popularity, link prediction, algorithms \\ \addlinespace
T01 & Conventional classification & learning, classification, features, detection, machine, machine learning, performance, task, accuracy, supervised, art, state art, training, state, framework, labeled, reviews, f1, problem, existing \\ \addlinespace
T02 & News and journalism & news, articles, news articles, article, outlets, sources, news outlets, fake, news sources, coverage, fake news, stories, news article, comments, misinformation, bias, source, readers, sharing, local news \\ \addlinespace
T03 & Language modeling and generative AI & llms, language, llm, ai, human, language llms, generated, gpt, tasks, fine, performance, text, reasoning, english, languages, framework, bias, biases, natural language, detection \\ \addlinespace
T04 & Pandemics and vaccination & covid, covid 19, 19, pandemic, 19 pandemic, vaccine, anti, misinformation, vaccination, vaccines, related, impact, health, stance, concerns, narratives, public, understand, population, conspiracy \\ \addlinespace
T05 & Politics and polarization & political, election, accounts, elections, polarization, discourse, party, politicians, partisan, campaigns, presidential, public, parties, leaning, engagement, right, political discourse, campaign, topics, candidates \\ \addlinespace
T06 & Sentiment and opinion mining & sentiment, topic, topics, text, sentiments, opinion, positive, negative, opinions, corpus, words, polarity, word, set, mood, semantic, emotions, related, mining, terms \\ \addlinespace
T07 & Communities and participation & communities, community, members, group, groups, participation, activity, subreddits, behavior, moderation, work, engagement, discussion, interactions, active, patterns, success, subreddit, dynamics, effect \\ \addlinespace
T08 & Events and collective attention & events, event, real, world, real world, time, real time, world events, messages, stream, response, summarization, event detection, crisis, temporal, streams, disaster, detection, short, relevant \\ \addlinespace
T09 & Identity and self-presentation & gender, profiles, people, profile, personality, self, age, friends, differences, personal, female, women, attributes, men, traits, male, likely, demographic, interactions, groups \\ \addlinespace
T10 & Search, Q\&A, and tagging & question, search, questions, answering, answer, answers, quality, question answering, tagging, tags, systems, queries, engine, search engine, community question, engines, search engines, knowledge, expertise, tag \\ \addlinespace
T11 & Health and social support & health, mental, mental health, support, posts, self, individuals, depression, emotional, public, post, risk, effects, findings, stress, public health, related, treatment, examine, discuss \\ \addlinespace
T12 & Hateful and abusive speech & hate, speech, hate speech, speech detection, detection, hateful, moderation, targets, target, attacks, comments, replies, toxic, offensive, understanding, harassment, automatic, language, conversations, toxicity \\ \addlinespace
T13 & Visual and multimodal media & videos, video, images, memes, visual, meme, image, multimodal, popularity, misinformation, channels, sharing, creators, popular, whatsapp, shared, channel, metadata, short, audio \\ \addlinespace
T14 & Location and mobility & location, mobility, urban, check, patterns, foursquare, city, locations, temporal, services, cities, spatial, places, geographic, human mobility, ins, check ins, behavior, time, mobile \\ \addlinespace
\end{longtable}
\endgroup

\section{Framing Codebook}
\label{app:codebook}
The definitions distinguish an intended research problem from the lexical rule used to approximate it. Example cues are illustrative; the regular expressions used in the analysis specify matching exactly. A cue can be incidental, and an absent cue does not establish an absent concern. Categories are not mutually exclusive.
\begingroup
\small
\begin{longtable}{@{}p{.20\textwidth}p{.46\textwidth}p{.26\textwidth}@{}}
\caption{Definitions and example lexical cues for the eight overlapping framing cues.}\label{tab:codebook}\\
\toprule
Category & Definition and scope limitation & Example cues \\
\midrule
\endfirsthead
\toprule
Category & Definition and scope limitation & Example cues \\
\midrule
\endhead
\bottomrule
\endfoot
P1 Structure and diffusion & Explicit framing of relational structure, diffusion, contagion, social influence, collective dynamics, or community formation. Does not identify every explanatory study of collective behavior. & diffusion; cascades; social ties; homophily \\ \addlinespace
P2 Prediction and classification & Explicit prediction, forecasting, classification, detection, or recommendation terminology. May match a prediction or detection task that a paper critiques rather than performs. & prediction; forecasting; classification; detection \\ \addlinespace
P3 Measurement and validity & Explicit validation, sampling, representation, reliability, ground truth, reproducibility, or measurement-error terminology. Routine uses of the verb measure alone do not qualify. Validation and reliability can refer to routine evaluation or background discussion, not a central measurement contribution. & validity; sampling; ground truth; reliability \\ \addlinespace
P4 Explanation and intervention & Explicit causation, intervention, recognizable experimental-design terminology, causal/social mechanisms, or theory testing. Generic mechanism and theoretical implication mentions are excluded. Theory, mechanism, or causal language does not establish a valid causal design. & causal; intervention; randomized; propensity score \\ \addlinespace
P5 Harm and information integrity & Explicit information-integrity, abuse, discrimination, privacy-risk, spam, or other harm framing. Includes descriptive and technical work about harm, without inferring the authors' normative commitments. & misinformation; spam; harassment; privacy; harm \\ \addlinespace
P6 Governance and participation & Explicit moderation, governance, censorship, enforcement, platform rules, or community norms. Does not capture every study of participation; moderation can also be mentioned as background. & moderation; governance; censorship; community rules \\ \addlinespace
P7 Support and well-being & Explicit human-health, care, distress, social-support, or well-being framing. The bare word health is excluded to reduce organizational-health metaphors. A health mention does not establish a clinical outcome or a supportive intervention. & mental health; caregiving; depression; social support \\ \addlinespace
P8 Resources and research access & Explicit nearby provision of data/tools or data-access and research-infrastructure terminology. Generic API and publicly available mentions alone do not qualify. A paper can provide a tool without belonging to a dataset or demonstration track. & researcher access; open-source; releases a dataset \\ \addlinespace
\end{longtable}
\endgroup

\section{Sensitivity Results}
\label{app:sensitivity}
Table~\ref{tab:sensitivity} reports the harm, community-governance, and platform-composition checks using the four main comparison periods.
Density is the mean of paper-level cue matches divided by title/abstract word count and multiplied by 100; it is not a paper prevalence. The historical union adds trolling, flaming, and vandalism terms to all editions. Spam is already in the baseline harm lexicon. The strict governance cue excludes papers whose only matches are \emph{moderate}, \emph{moderated}, \emph{moderates}, or \emph{moderating}.
\begingroup
\small
\begin{longtable}{@{}p{.36\textwidth}rrrr@{}}
\caption{Sensitivity estimates. All rows except cue density are percentages.}\label{tab:sensitivity}\\
\toprule
Measure & 2007--11 & 2012--16 & 2017--21 & 2022--26 \\
\midrule
\endfirsthead
\toprule
Measure & 2007--11 & 2012--16 & 2017--21 & 2022--26 \\
\midrule
\endhead
\bottomrule
\endfoot
Harm: full title/abstract & 6.2 & 12.3 & 27.0 & 38.1 \\ \addlinespace
Harm: title + first 100 abstract words & 5.7 & 9.5 & 22.7 & 31.9 \\ \addlinespace
Harm: first 100 abstract words only & 5.7 & 8.7 & 21.9 & 31.0 \\ \addlinespace
Harm: historical union & 6.2 & 12.3 & 27.9 & 38.3 \\ \addlinespace
Harm: mean matches / 100 words & 0.10 & 0.23 & 0.54 & 0.78 \\ \addlinespace
Governance within community topic & 4.0 & 1.2 & 14.4 & 34.3 \\ \addlinespace
Governance after feature masking & 3.6 & 1.1 & 13.4 & 32.5 \\ \addlinespace
Governance within community: title + first 100 words & 2.5 & 0.8 & 8.8 & 25.8 \\ \addlinespace
Governance within community: first 100 abstract words & 2.5 & 0.8 & 7.4 & 25.6 \\ \addlinespace
Governance within community: strict cue & 2.5 & 0.5 & 12.5 & 33.3 \\ \addlinespace
Governance cues overall: strict cue & 0.8 & 0.6 & 5.5 & 13.1 \\ \addlinespace
Governance within community: platform-standardized & 4.0 & 1.6 & 11.6 & 29.7 \\ \addlinespace
Governance within community: no Twitter/Reddit mention & 4.3 & 1.8 & 11.6 & 30.1 \\ \addlinespace
\end{longtable}
\endgroup

\subsection{Platform composition}
Platform contexts change across the corpus, and Reddit research is comparatively governance-heavy, so the community--governance contrast could reflect which platforms the conference studied. 
We use the frozen Reddit and Twitter indicators to classify each contribution as mentioning Reddit, mentioning Twitter without Reddit, or mentioning neither, and recompute the conditional rate within strata. Standardizing every period to the 2007--2011 stratum distribution of community-topic weight gives 4.0, 1.6, 11.6, and 29.7\%, an endpoint difference of 25.7 points [15.2, 35.6] against the crude 30.3 points [21.8, 38.4]. Among contributions mentioning neither platform---93\% of community-topic weight in the first period and 44\% in the last---the rate moves from 4.3 to 30.1\%, a difference of 25.8 points [14.3, 36.5]. Platform turnover therefore accounts for about one sixth of the contrast.
\subsection{Direct term overlap}
Table~\ref{tab:overlap} gives the number of each topic's top twenty terms matched by each framing expression. For the community component, only \emph{moderation} matches the governance expression. The vocabulary-removal analysis masks every governance-matching unigram or bigram in the fitted vocabulary, not just the top twenty terms. The complete term-level overlap record will be released with the analysis materials upon publication.
\begingroup
\small
\begin{longtable}{@{}lrrrrrrrr@{}}
\caption{Counts of top-twenty topic terms matched by each framing lexicon. These are literal term matches, not measures of semantic independence.}\label{tab:overlap}\\
\toprule
Topic & P1 & P2 & P3 & P4 & P5 & P6 & P7 & P8 \\
\midrule
\endfirsthead
\toprule
Topic & P1 & P2 & P3 & P4 & P5 & P6 & P7 & P8 \\
\midrule
\endhead
\bottomrule
\endfoot
T00 & 2 & 3 & 0 & 0 & 0 & 0 & 0 & 0 \\ \addlinespace
T01 & 0 & 2 & 0 & 0 & 0 & 0 & 0 & 0 \\ \addlinespace
T02 & 0 & 0 & 0 & 0 & 2 & 0 & 0 & 0 \\ \addlinespace
T03 & 0 & 1 & 0 & 0 & 0 & 0 & 0 & 0 \\ \addlinespace
T04 & 0 & 0 & 0 & 0 & 2 & 0 & 7 & 0 \\ \addlinespace
T05 & 0 & 0 & 0 & 0 & 0 & 0 & 0 & 0 \\ \addlinespace
T06 & 0 & 0 & 0 & 0 & 0 & 0 & 0 & 0 \\ \addlinespace
T07 & 0 & 0 & 0 & 0 & 0 & 1 & 0 & 0 \\ \addlinespace
T08 & 0 & 2 & 0 & 0 & 0 & 0 & 0 & 0 \\ \addlinespace
T09 & 0 & 0 & 0 & 0 & 0 & 0 & 0 & 0 \\ \addlinespace
T10 & 0 & 0 & 0 & 0 & 0 & 0 & 0 & 0 \\ \addlinespace
T11 & 0 & 0 & 0 & 0 & 0 & 0 & 3 & 0 \\ \addlinespace
T12 & 0 & 2 & 0 & 0 & 5 & 1 & 0 & 0 \\ \addlinespace
T13 & 0 & 0 & 0 & 0 & 1 & 0 & 0 & 0 \\ \addlinespace
T14 & 0 & 0 & 0 & 0 & 0 & 0 & 0 & 0 \\ \addlinespace
\end{longtable}
\endgroup

\section{Exploratory Temporal Segmentation}
\label{app:boundaries}
For segmentation only, we normalize annual overlapping framing-cue rates together with an unmatched category. We concatenate square-root topic and framing shares, weighting the axes equally, and minimize within-segment squared deviation. Each edition has equal weight in this diagnostic. This differs from pooling papers within the fixed periods used in the main analysis.

The baseline five-segment solution with a three-edition minimum has starts in 2007, 2011, 2017, 2021, and 2024. Its full partition recurs in 29/200 resamples (14.5\%). Exact boundary frequencies are given below. The baseline minimum makes 2024 the latest possible final start, so its apparent recurrence cannot establish a recent historical break. A second analysis uses the same 200 newly sampled paper compositions to compare two- and three-edition minima. It admits a 2025 start, which appears in 47/200 resamples (23.5\%). These are conditional exact-year frequencies, not confidence levels for historical events.
\begingroup
\small
\begin{longtable}{@{}lrrr@{}}
\caption{Exact-year boundary recurrence (percent) under five-segment joint profiles. A dash marks a boundary disallowed by the minimum-length rule. The two new columns use identical resampled paper compositions.}\label{tab:boundaries}\\
\toprule
Boundary start & Baseline 200; min. 3 & New 200; min. 3 & Same new 200; min. 2 \\
\midrule
\endfirsthead
\toprule
Boundary start & Baseline 200; min. 3 & New 200; min. 3 & Same new 200; min. 2 \\
\midrule
\endhead
\bottomrule
\endfoot
2009 & -- & -- & 21.5 \\ \addlinespace
2010 & 39.5 & 39.0 & 33.0 \\ \addlinespace
2011 & 56.5 & 59.0 & 54.5 \\ \addlinespace
2012 & 3.5 & 1.5 & 28.0 \\ \addlinespace
2013 & 19.5 & 20.5 & 5.0 \\ \addlinespace
2014 & 16.5 & 13.0 & 8.0 \\ \addlinespace
2015 & 6.5 & 9.0 & 5.5 \\ \addlinespace
2016 & 1.0 & 2.5 & 0.5 \\ \addlinespace
2017 & 54.0 & 53.0 & 55.5 \\ \addlinespace
2018 & 34.0 & 34.0 & 31.5 \\ \addlinespace
2019 & 11.5 & 12.5 & 23.0 \\ \addlinespace
2020 & 21.5 & 19.5 & 14.5 \\ \addlinespace
2021 & 65.0 & 65.5 & 70.5 \\ \addlinespace
2022 & 7.0 & 5.5 & 4.0 \\ \addlinespace
2023 & 0.0 & 0.0 & 0.0 \\ \addlinespace
2024 & 64.0 & 65.5 & 21.5 \\ \addlinespace
2025 & -- & -- & 23.5 \\ \addlinespace
\end{longtable}
\endgroup

We also retain results for three, four, and five segments and topic-only, framing-only, and joint representations. The segment count is not selected by an external validation criterion. Changes in component count and initialization can move boundaries. We therefore do not name a definitive sequence of five phases or use it to calculate the headline period comparisons. Broad emphases can overlap even when a segmentation algorithm must assign each year to one interval.

\section{Independent Validation Protocol}
\label{app:validation}
The framing-validation protocol samples 20 papers per edition (400 total), without restricting to cue-positive cases. Two coders independently judge whether each abstract expresses a substantive objective matching each category, distinguishing central or evaluated concerns from incidental background. Multiple categories are allowed; uncertain cases receive a separate flag. The protocol materials withhold automatic predictions. Source links can reveal bibliographic identity; an optional local reader also withholds author and edition information.

Report per-category Cohen's $\kappa$, raw agreement, and positive-label counts overall and by fixed period. After preserving the independent ratings, adjudicate disagreements and uncertainty. Compare the frozen cues with those adjudicated judgments, reporting precision, recall, and the TP/FP/FN/TN counts for each category and period. Since each edition contributes twenty coded papers regardless of its size, pooled period estimates should use inverse-inclusion weights $N_t/20$. Report denominators and unresolved judgments; undefined recall for a category with no positive gold labels must not be replaced by zero. Development-sample overlaps are flagged for a separate sensitivity analysis. Any lexicon revision based on these judgments requires a new held-out evaluation.

For topic interpretability, the protocol contains five word-intrusion items per component (75 total). Each presents five high-weight terms and an intruder drawn from another component, with an IDF-matching rule to reduce rarity cues. Eighty topic-intrusion items (four papers per edition) present the three highest-weight topic word lists and one low-weight list for the document. Coders select the intruder without seeing the model weights or key. Report correct-choice rates, counts, uncertainty, and comments by component or period; chance rates are $1/6$ and $1/4$, respectively. These tasks assess interpretability; they do not independently establish the substantive validity of a label.

\end{document}